# Phase evolution of Peregrine-like breathers in optics and hydrodynamics

**Gang Xu [1], Kamal Hammani [1], Amin Chabchoub [2], John M. Dudley [3], Bertrand Kibler [1], and Christophe Finot [1,*]**

[1] *Laboratoire Interdisciplinaire Carnot de Bourgogne, UMR 6303 CNRS-Université de Bourgogne-Franche-Comté, 9 avenue Alain Savary, BP 47870, 21078 Dijon Cedex, France*

[2] *Centre for Wind, Waves and Water, School of Civil Engineering, The University of Sydney, Sydney, New South Wales 2006, Australia*

[3] *Institut FEMTO-ST, CNRS Université de Bourgogne Franche-Comté UMR 6174, 25030 Besançon, France*

[*] *Corresponding author:*
E-mail address: *christophe.finot@u-bourgogne.fr*
*Tel.: +33 3 80395926*

**Abstract:** We present a detailed study of the phase properties of rational breather waves observed in the hydrodynamic and optical domains, namely the Peregrine soliton and related second-order solution. At the point of maximum compression, our experimental results recorded in a wave tank or using an optical fiber platform reveal a characteristic phase shift that is multiple of $\pi$ between the central part of the pulse and the continuous background, in agreement with analytical and numerical predictions. We also stress the existence of a large longitudinal phase shift across the point of maximum compression.

# 1. Introduction

The nonlinear Schrödinger equation (NLSE) is a rather simple but extremely powerful generic equation, able to describe the evolution of a large range of nonlinear waves, including hydrodynamic waves, Bose-Einstein condensates or light in single-mode fibers. Among the existing solutions that can be analytically computed, bright solitons are undoubtedly the best known, having been deeply studied in various fields since the early 70s [1]. Solitons on finite background (also called breathers) have also been the subject of analytical investigations in the early 80s with nonlinear structures, now known as the Kuznetsov-Ma solitons [2], the Peregrine soliton (PS) [3] and the Akhmediev breathers (AB) [4].

In the last decade, these nonlinear coherent structures presenting temporal and/or spatial localization have been extensively investigated after the renewed interest driven primarily by the study of extreme events [5]. This has led to the first experimental observation of breather waves with typical signatures of the Peregrine soliton, both in the optical and hydrodynamic domains [6, 7] as well as in multicomponent plasmas [8]. Those experiments have confirmed the temporal and spectral exchange of energy between the central localized peak and the finite background. With accurate generation and detection devices, the longitudinal evolutions of the temporal and spectral intensity profiles have been carefully characterized and found to be in good agreement with the analytical predictions. Since these pioneering experiments, many other works have confirmed the crucial importance of PS or AB features in the understanding of rogue events induced by modulation instability [9-11], in the higher-order soliton compression stage [12], in the focusing evolution of super-Gaussian structures [13, 14], as well as in the turbulent evolution of a partially incoherent wave [15-18]. At the same time, higher-order rational solutions have also been identified [5, 19] and experimentally synthesized up to the fifth order [20-22].

If examples confirming the temporal and spectral profiles of the PS are numerous, much less attention has been devoted to the characterization of their phase profile. To the best of our knowledge, no results are up to now available in the hydrodynamic domain and only few works exist in fiber optics [6, 12, 23], stressing the existence at the point of maximum compression of a typical $\pi$ phase shift between the central peak of the PS and the continuous background.

In this paper, we specifically focus our analysis on the phase properties of the PS both in time and space. We revisit some of our past experimental works and provide new results to emphasize the phase shifts that is experienced at the point of maximum compression, for rational breather solutions, namely the Peregrine soliton and related higher-order solution. The paper is therefore organized as follows. First, we recall the analytical properties of ideal PS and the second order rational soliton. Then, we investigate the phase profile of such nonlinear waves experimentally recorded at the point of maximum focusing in a water tank. In a third section, we focus our attention to fiber optics and we also confirm the existence of a π-phase shift at the point of maximum compression between the central peak and the surrounding background. In addition, we also follow the longitudinal evolution of this phase difference along propagation, thus providing full characterization of the PS growth-decay cycle.

## 2. Principle under investigation and properties of rational solitons

As stressed by several recent studies [24, 25], the propagation of weakly-nonlinear water waves in one-dimensional system or light-waves in single-mode optical fibers can be described by the focusing nonlinear Schrödinger equation written in the following dimensionless form :

$$i\psi_\xi + \frac{1}{2}\psi_{\tau\tau} + |\psi|^2 \psi = 0 \qquad (1)$$

where subscripted variables stand for partial differentiations. Here $\psi$ is a wave group or wave envelope which is a function of $\xi$ (a scaled propagation distance or longitudinal variable) and $\tau$ (a comoving time, or transverse variable, moving with the velocity of the wave-group central frequency).

More precisely, in the case of hydrodynamics, $\psi$ is related to the water wave elevation $\eta(z,t)$ to the lowest order by $\eta(z,t) = \text{Re}\{\psi(z,t) \exp(i(k_0 z - \omega_0 t))\}$ with $k_0$ being the wave number of the carrier wave and $\omega_0$ its angular frequency given by the dispersion relation of linear deep-wave theory, $\omega_0 = (g\, k_0)^{1/2}$, $g$ being the gravitational acceleration. The dimensional distance $z$ and time $t$ are related to the rescaled variables $\tau$ and $\xi$ by $z = \tau/\left(\sqrt{2}\, k_0^2\, a_0\right) + c_g\, t$ and $t = 2\,\xi/\left(k_0^2\, a_0^2\, \omega_0\right)$ with $c_g = \omega_0/(2k_0)$ being the group velocity and $a_0$ being the initial amplitude of the carrier wave.

In the case of optics, the normalized quantity $\psi$ is related to the complex slowly-varying envelope $A(t,z)$ of the electrical field by $A = P_0^{1/2} \psi$, $P_0$ being the average power assuming a perturbed continuous wave excitation. The dimensional distance $z$ and time are related to the previous normalized parameters by $z = \xi L_{nl}$ and $t = \tau t_0$ where the characteristic nonlinear length and time scales are $L_{nl} = 1/(\gamma P_0)$ and $t_0 = (|\beta_2| L_{nl})^{1/2}$, respectively. $\gamma$ is the nonlinear coefficient of the fiber and $\beta_2$ its second-order dispersion coefficient.

The solution derived by H. Peregrine has a particular fractional form that has led this class of solution to be described as 'rational soliton'. The first-order rational soliton is given by [3]:

$$\psi_1(\xi,\tau) = \left[1 - \frac{4(1+2i\xi)}{1+4\tau^2+4\xi^2}\right] e^{i\xi} \qquad (2)$$

where $\xi = 0$ corresponds to the point of maximum compression of the breathing structure. The modulus of the spectral intensity profile $|\psi(\xi,\omega)|$, can also be mathematically expressed as (with the constant continuous background here omitted) [26]:

$$|\psi_1(\xi,\omega)| = \sqrt{2\pi} \exp\left(-\frac{|\omega|}{2}\sqrt{1+4\xi^2}\right), \qquad (3)$$

leading to a characteristic triangular shape when plotted on a logarithmic scale. Experimentally speaking, according to the domain of investigation, various quantities can be easily recorded. The field $\psi(\xi,\tau)$ can be recovered from the water elevation at a given position. However, in optics, for picosecond events, it is easier to have access to the temporal and spectral intensity profiles $|\psi(\xi,\tau)|^2$ and $|\psi(\xi,\omega)|^2$.

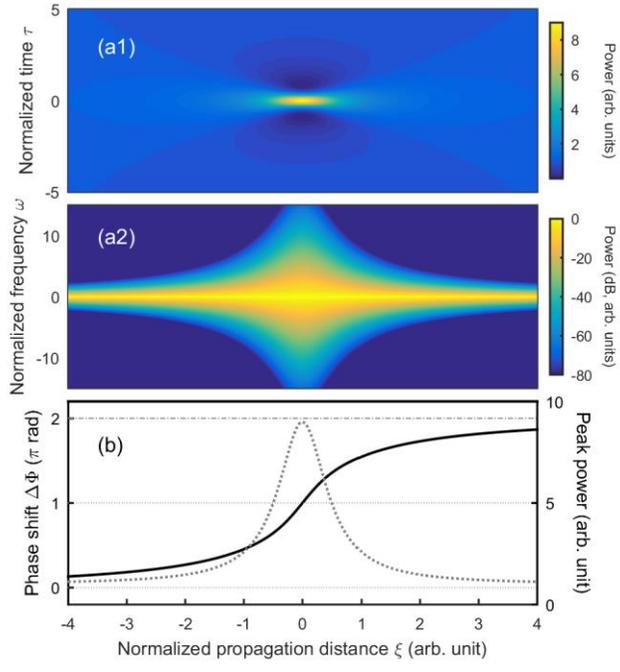

**Figure 1:** Longitudinal evolution of the ideal Peregrine soliton: (a) Temporal and spectral power profile $|\psi|^2$ (panels (a1) and (a2) respectively). (b) Evolution of the phase shift between the central part of the continuous background $\Delta\phi$. (left, solid black line, for clarity of the figure, the phase has been unwrapped) and peak intensity (right, dot dashed line).

The longitudinal evolution of $|\psi(\xi,\tau)|^2$ is displayed in Fig. 1(a). The PS is a limiting case of the $\tau$-periodic AB and the $\xi$-periodic Kuznetsov-Ma breather when the period tends to infinity. Therefore, PS is doubly localized and the maximal amplification, obtained at $\xi = \tau = 0$ is 9 times the background intensity. The temporal field and power profiles at the point of maximum compression (corresponding also to the point of maximum spectral extension) are shown in Fig. 2 (panels (a1) and (b1) respectively, left column). At $\tau = 0$, the field is real and experiences a sign inversion between the central peak and the continuous wave background. This corresponds to a $\pi$ phase shift $\Delta\phi$ between the two temporal parts of the pulse as can be seen in panel Fig. 2 (c1) representing the phase profile $\varphi(0,\tau) = arg(\psi(0,\tau))$. $\Delta\phi$ is here defined as $\Delta\phi(\xi) = \varphi(\xi,\tau=0) - \varphi(\xi, \tau \to \infty)$. The longitudinal evolution of $\Delta\phi$ can be easily derived analytically:

$$\tan(\Delta\phi) = -\frac{8\,\xi}{4\,\xi^2 - 3} \qquad (4)$$

and is plotted in panel (b) of Fig. 1. For the ideal PS, a phase excursion of $2\pi$ does exist, as already noticed in works discussing properties of Fermi-Pasti-Ulam recurrence [27]: after a full growth-decay cycle, the initial and final states are identical, but with a $2\pi$ phase shift accumulated during the nonlinear recurrence cycle.

Higher-order rational solitons on finite background can also be observed. For the second-order rational solution, the analytical expression becomes more complex and can be obtained from Darboux transformation [5]:

$$\psi_2(\xi,\tau) = \left[1 + \frac{G + i\,H}{D}\right] e^{i\xi} \tag{5}$$

with $G$, $H$ and $D$ being the polynomial functions given by:

$$\begin{aligned}
G(\xi,\tau) &= \frac{3}{16} - \frac{3}{2}\tau^2 - \tau^4 - \frac{9}{2}\xi^2 - 6\,\tau^2\,\xi^2 - 5\,\xi^4 \\
H(\xi,\tau) &= \left(\frac{15}{8} + 3\,\tau^2 - 2\,\tau^4 - \xi^2 - 4\,\tau^2\,\xi^2 - 2\,\xi^4\right)\xi \\
D(\xi,\tau) &= \frac{3}{64} + \frac{9}{16}\tau^2 + \frac{1}{4}\tau^4 + \frac{1}{3}\tau^6 + \frac{33}{16}\xi^2 \\
&\quad - \frac{3}{2}\tau^2\,\xi^2 + \tau^4\,\xi^2 + \tau^2\,\xi^4 + \frac{9}{4}\xi^4 + \frac{1}{3}\xi^6
\end{aligned} \tag{6}$$

Profiles of the second-order rational solution is provided in panels a2-c2 of Fig. 2 (right column). The peak power of this structure is significantly higher than the PS: as the amplitude of the $n^{th}$ order structure wave scales with a factor ($2n+1$), the peak intensity is now boosted by a factor 25 with respect to the continuous background. In this case, the structure crosses the zero value four times so that the central peak is now in phase with the continuous background.

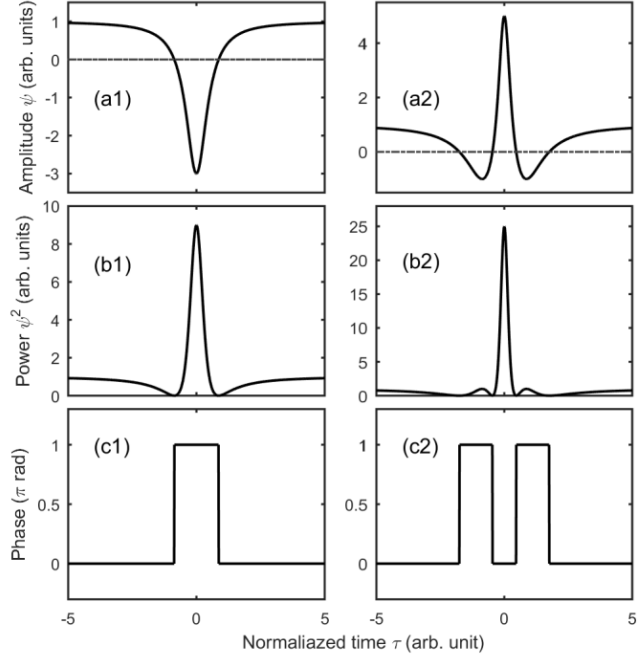

**Figure 2:** Temporal amplitude $\psi$, intensity $\psi^2$ and phase profiles $\varphi$ (panels a, b and c respectively) at the point of maximum compression for a fundamental Peregrine soliton $\psi_1$ and for a 2$^{nd}$ rational soliton $\psi_2$ (panels 1, 2 respectively).

## 3. Phase profiles for hydrodynamic waves

We first discuss the results obtained for water waves, in particular features of the PS and the 2$^{nd}$-order rational soliton observed at the point of maximum compression.

### 3.1 Experimental setup

The experimental setup is described in details in Refs. [20, 28]. The experiments have been performed in the Hamburg University of Technology in a $15 \times 1.6 \times 1.5$ m$^3$ water wave tank with 1 m water depth. A single-flap wave-producing paddle activated by a hydraulic cylinder is located at the far end of the tank. The assumption that the flap displacement is proportional to the generated surface height has been verified by measurements. In order to avoid higher-order nonlinear effects, we have kept the steepness $a_0\,k_0$ of the wave very low [28]. As next, we will study two different breather wave evolutions. We first studied the fundamental Peregrine soliton with $a_0 = 1$ cm and $\omega_0 = 10.7$ rad/s, leading to a steepness of 0.12 [28]. We also synthetized the second-order rational soliton with $a_0 = 1$ mm and $\omega_0 = 17.2$ rad/s leading to a steepness of 0.03 [20]. To avoid wave reflections, an absorbing beach is installed at the opposite end. All experiments are conducted under conditions where the ratio of the water depth of 1 m to the wavelength is larger than unity. The surface height of the water at a given position is measured by a capacitance wave gauge with a sensitivity of 1.06 V/cm which allows us to make measurements with an accuracy of up to three significant digits. The sampling frequency is 500 Hz. The field $\psi$ is extracted from the experimental measurements $\eta$ using a Hilbert transform [29, 30] after spectral filtering to isolate the fundamental component of the spectra. Note that in order to increase the effective propagation length for the observation of the 2$^{nd}$ order rational soliton, we split the experiment into several stages. Namely, starting the wave generation repetitively with different boundary conditions given from theory, we measured the wave profile at the other end of the tank. We repeat this process 7 times, thus reaching the propagation distance of 72 m, which corresponds to the point of maximum wave amplitude.

## 3.2 Experimental results

Results obtained for the fundamental PS are summarized in Fig. 3 and are compared with the analytical envelope profile predicted by Eq. (1). The overall waveform is in excellent agreement with the analytical predictions. As expected, the maximum amplitude of the envelope is three times as high as the amplitude of the constant background, and at that moment, the central localized structure crosses the zero value twice. Remarkably, the results of the phase reconstruction (panel (c)) highlights a clear π-phase shift at the point of maximum compression between the central part and the background wave, as expected by the analytics: the pulsed part of the wave is in antiphase with respect to the carrier wave.

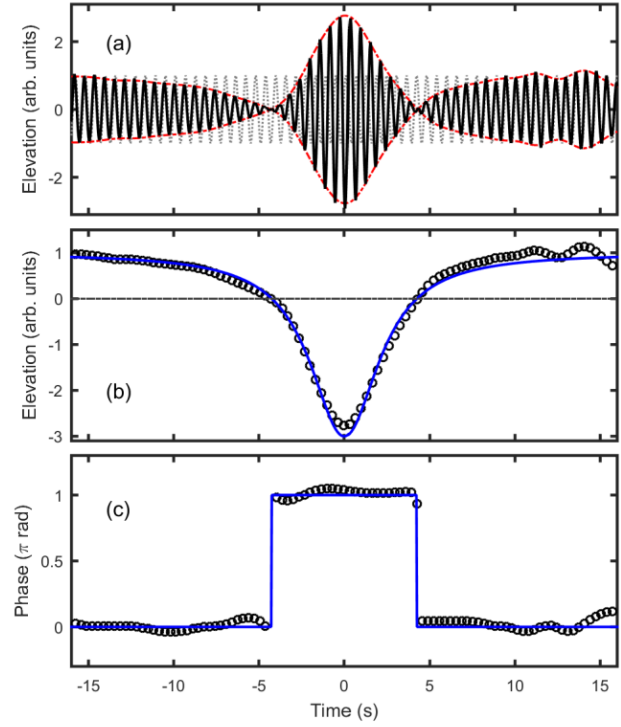

**Figure 3:** Profile of the fundamental PS at the point of maximum compression. (a) Recorded surface elevation (black curve) with corresponding wave envelope (red curve) and carrier wave as visual reference (grey line). Experimental envelope $\psi$ and phase profile $\varphi$ (black circles in panel (b) and (c) respectively) are compared with the analytical predictions (solid blue curve) provided by Eq. (2).

A second series of measurements have dealt with the generation of second-order rational structure. Results recorded at the point of maximum compression are summarized in Fig 4. As expected, the nonlinear focusing of the water wave leads to an amplification by a factor 5 of the amplitude of background value, in agreement with the analytical form. The envelope presents four passages through zero value, confirming that a $n^{th}$ order rational soliton presents $2n$ zeros [31]. We observe that, according to the position within the nonlinear structure, phase shifts of multiples of $\pi$ can be observed. Despite some distortions, those typical phase shifts are in agreement with the expected results from theory. The central part is in this case in phase with the continuous background.

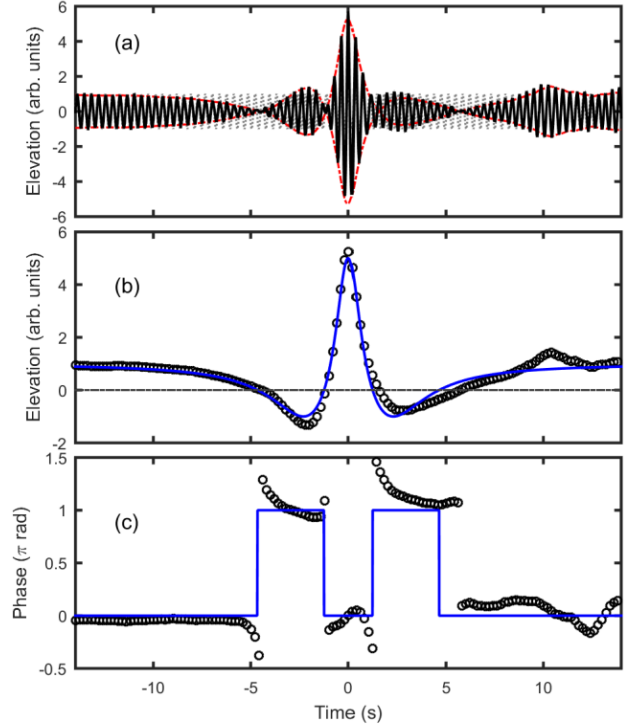

**Figure 4:** Profile of the second-order rational soliton at the point of maximum compression. (a) Recorded surface elevation (black curve) with corresponding wave envelope (red curve) and carrier wave as visual reference (grey line). Experimental envelope $\psi$ and phase profile (black circles in panel (b) and (c) respectively) are compared with the analytical predictions (solid blue curve) provided by Eq. (5) and (6).

## 4. Phase evolution in optical fiber

In this second part, we now focus on the nonlinear propagation of light waves in a single-mode fiber.

### 4.1 Optical setup

The experimental setup is based on commercially available equipment of the telecom industry. In previous experiments in optics, we used the sinusoidal beating resulting from the temporal interference of two frequency-offset continuous waves (CW) [6] or a CW laser externally modulated by a simple Lithium Niobate intensity modulator (IM) driven by a sinusoidal modulation [32]. Despite their ease of implementation, those configurations have proven very efficient to demonstrate the main features of PS or AB, the sinusoidal wave evolving progressively towards the targeted nonlinear structure [33]. However, one restriction due to the imperfect initial excitation is that after the first stage of growth, the return to the initial stage is only partial and the nonlinear structure tends to split into several higher-order structures [32, 34]. For an efficient reshaping, the input sinusoidal modulation should be rather small, therefore requiring significant propagation distance. The accumulated losses may ultimately affect the growth and decay cycle. In order to avoid those various restrictions [6], we implemented a more advanced configuration which is described in Fig. 5. A frequency comb with a 20-GHz line spacing is first generated by the nonlinear evolution of a sinusoidal beating in a fiber. In order to have input conditions as close as possible to the ideal PS, the discrete spectral components are then spectrally shaped in amplitude as well as in phase using a liquid crystal on silicon based programmable filter (Waveshaper device) [21, 35]. This method enables us to synthetize as an initial condition a close-to-ideal PS wave at any propagation length. Let us note that, given the temporal periodicity of the input wave, we cannot generate rigorously a single PS, but rather a train of breather waves where each individual element fits the limiting ideal case of PS (both in amplitude and phase). The corresponding IST spectrum would preserve the global information of the PS [36]. Note that a phase modulator (PM) is also inserted in the initial comb source, in order to prevent the deleterious effects of Brillouin backscattering. Next the resulting shaped wave is amplified by a high-power erbium doped fiber amplifier (EDFA) up to average powers of 28.5 dBm. The propagation takes place in a combination of segments of variable lengths made of the most standard fiber that is currently available, i.e. the single-mode fiber SMF-28 that is compliant with International Telecommunication Union recommendation G.652. The anomalous group-velocity dispersion of this fiber is $\beta_2 = -21$ ps$^2$/km whereas its nonlinear coefficient $\gamma$ is 1.1 /W/km, leading to a nonlinear length $L_{nl}$ of 1260 m. Given the high value of $\beta_2$, the impact of third order dispersion is negligible in the spectral bandwidth under study. After propagation into the fiber, the output field is recorded in the temporal domain taking advantage of an optical sampling oscilloscope (Picosolve PSO100 series) that allows a temporal resolution of the order of the picosecond. The output spectral properties are also recorded using an optical spectrum analyzer with a spectral resolution of 2.5 GHz (Yokogawa AQ6370). Note that given the high repetition rate that is here considered, it is not possible to involve signal real-time characterization schemes, such as dispersion Fourier transform or temporal lens [37]. Although previous phase reconstruction of the output profiles have benefited from Frequency Resolved Optical Gating [6] or heterodyne approaches [23], we have used in this work the Gerchberg-Saxton algorithm [38] that only requires the knowledge of the temporal and spectral intensity profiles and which has previously been used in recent experiments on laser pulse characterization [37]. Considering the spectral comb nature of the signal greatly simplifies the convergence of the algorithm, with only a few tens of components having to be involved. The phase and time orientation ambiguity that may intrinsically exist in this algorithm is removed by comparison with realistic numerical simulation of the NLSE. As the GS algorithm cannot provide the absolute phase of optical wave, we have fixed arbitrarily the phase of the background to zero value.

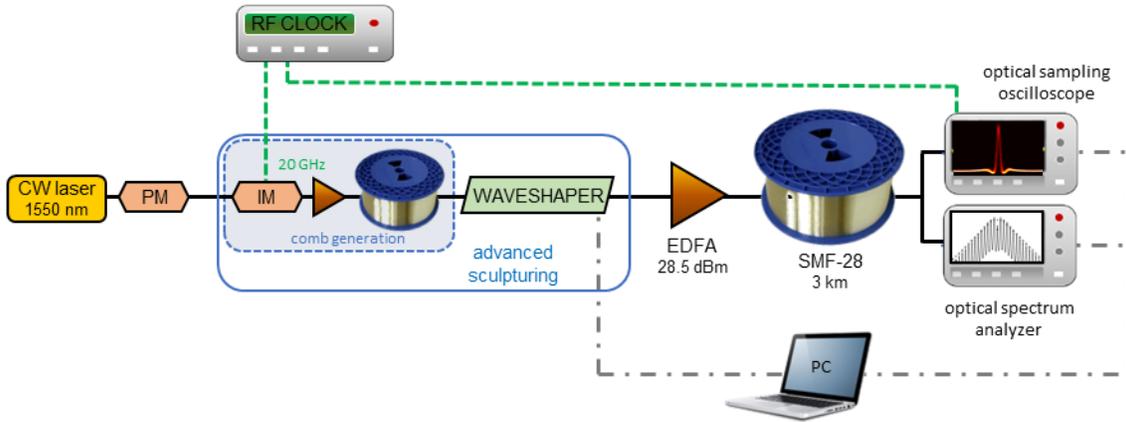

**Figure 5:** NLS breather generator based on light wave propagation in a single-mode optical fiber. PM: phase modulator; IM: intensity modulator; EDFA: Erbium-doped fiber amplifier.

## 4.2 Experimental results

We first investigate the longitudinal evolution of the temporal and spectral intensity profiles. The input profile programmed on the spectral waveshaper corresponds to the PS-like structure at a normalized distance $\xi = z / L_{nl} = -1.2$. We carried out 22 measurements, involving fiber lengths up to 3 km, corresponding to a normalized length from $\xi = -1.2$ to $\xi = 1.2$. Experimental results are summarized in Fig. 6 and reproduce the spatiotemporal localization similar to the ideal Peregrine wave (panels 2). The point of maximum temporal compression occurs after 1.5 km of nonlinear propagation. This distance also corresponds to the maximal spectral extent of the optical structure (see panel b). It is worth mentioning that contrary to previous experimental realization based on approximate sinusoidal inputs, the recorded longitudinal evolution is here rather symmetric and does not present any sign of pulse splitting. The experimental measurements are in excellent agreement with the analytical evolution of an ideal PS (panels 2).

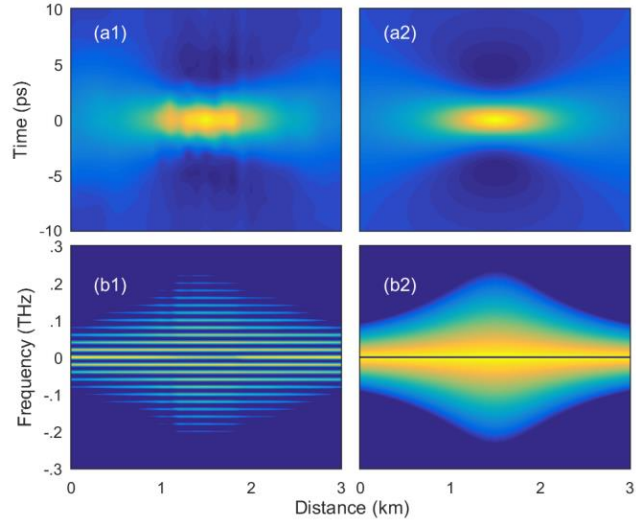

**Figure 6 :** Longitudinal evolutions of the temporal (a) and spectral (b) intensity profiles of a fundamental PS. Experimental results (panels 1) are compared with ideal PS predicted from Eq. (2) and (3) and using dimensional units of experiments (panels 2). The results of the temporal intensity profiles are plotted on a linear scale whereas the spectra as displayed using a logarithmic colormap with a 33 dB dynamics.

Details of the temporal phase and intensity profiles at the point of maximum temporal compression are provided in Fig. 7 (panels (a) and (b)) and plotted over a temporal window corresponding to one period of the initial 20-GHz excitation. The temporal profile retrieved at a distance z = 1.5 km exhibits the typical signatures of the PS. Compared to the initial localized perturbation (red line) obtained after accurate phase and amplitude sculpturing of the frequency comb, the wave has been significantly compressed down to a full-width at half-maximum of 3 ps. The ratio between the background and the central peak is up to 8. The sharp phase shift between the central part and the continuous background has also increased up to a value that becomes close to π. Therefore, the reconstructed field passes twice through the zero value and is in convincing agreement with the typical the PS profile corresponding to the parameters involved in the experiment (no fitted procedure has been here employed).

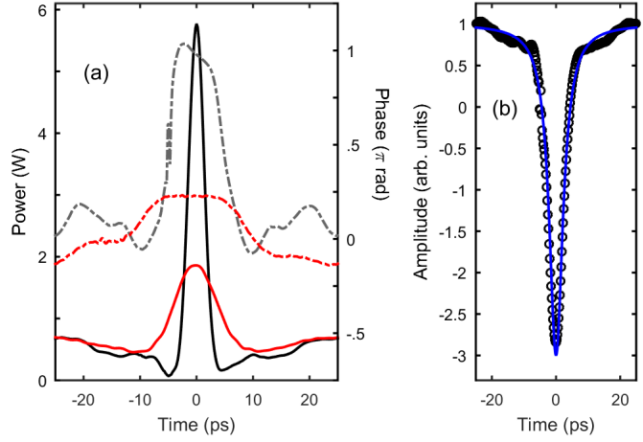

**Figure 7: (a)** Experimental intensity and phase profiles obtained for the generated Peregrine-like soliton (solid and dashed-dotted lines respectively). The experimental results retrieved at the point of maximum compression (black and grey lines) are compared with the synthetized waveform used as the initial perturbation condition (red curve). The phase of the background is here arbitrarily fixed to 0. **(b)** Real part of the amplitude profile at the point of maximum compression: experimental results (circles) are compared with analytical shape of the corresponding PS (circles).

In Fig. 8, we have also plotted the temporal profiles of the pulse at two distances symmetrically located with respect to the point of maximum compression, i.e, at distance z = 1000 and 2000 m (i.e $\xi$ = -0.4 or 0.4). In those two cases, the phase offset between the central part is not π anymore. However, we can note that, despite the sign of the central phase jump, the two experimental waveforms are extremely close, confirming the phenomenon of "déjà vu" and the Fermi-Pasta-Ulam recurrence of breathing structures [39]. Such an observation is in full agreement with the ideal PS evolution.

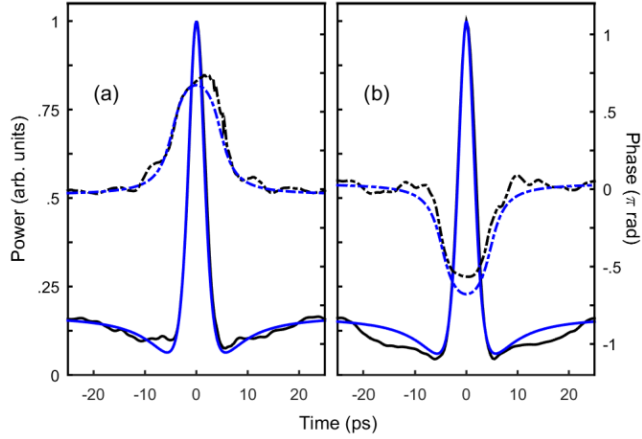

**Figure 8:** Different intensity and phase temporal profiles (solid and dashed-dotted lines respectively) symmetrically located before and after the point of maximum compression (distance of 1000 and 2000 m, panel a and b respectively). Experimental retrieved results (black lines) are compared with the ideal PS shape (blue line). The phase of the background is here arbitrarily fixed to 0.

A more systematic study of the magnitude of $\Delta\phi(z)$ retrieved by the Gerchberg-Saxton algorithm is shown in Fig. 9. We clearly see that the wave exhibits a longitudinal evolution of the phase difference between the wings and central region that agrees with the analytical predictions of an ideal PS (panel a). For the distances that have been here experimentally considered where attenuation is not significant, the phase excursion of $\Delta\phi$ is higher than $\pi$. In theory, starting from $\xi \rightarrow -\infty$, it can reach up to $2\pi$ at $\xi \rightarrow \infty$ [27]. A convenient way to observe the growth and decay cycle is to use a polar plot of the complex field at the point $\tau = 0$ (panel b) [27]. For an ideal PS, the trajectory on the complex plane describes a circle of radius 2 centered in {-1,0}. Once again, we see that our experiments qualitatively capture the main features of the evolution of an ideal PS over one whole cycle. The slight deviations that may be observed may be explained by several factors that encompass the imperfect excitation (we involve a periodic train of PS-like breathers instead of a single ideal one) as well as the residual dissipation of the fiber. Those limiting factors will especially impact the propagation for longer distances and may lead to phase-shifted spatial recurrence [40, 41].

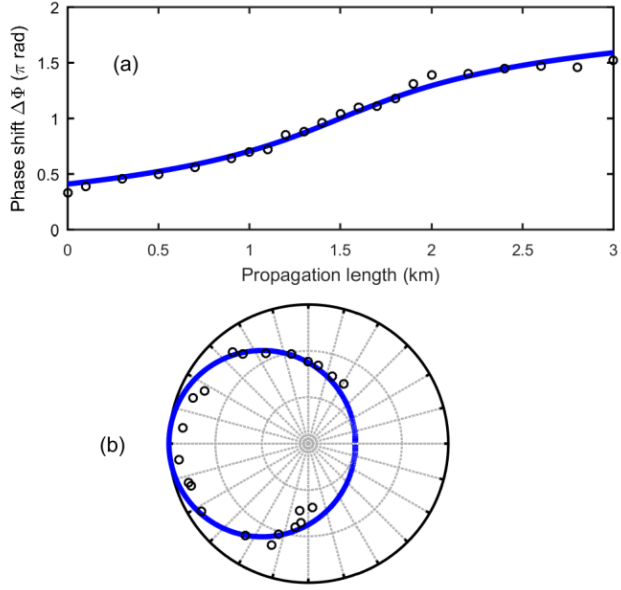

**Figure 9: (a)** Longitudinal evolution of the phase offset between the pulse central peak and the continuous background $\Delta\phi$. Experimental results (circles) are compared with analytical results based on Eq. (4) (blue line). For clarity, the phase has been unwrapped. **(b)** Polar representation of the longitudinal evolution of the complex field recorded at its maximum temporal value (normalized to the continuous background). Experimental results (circles) are compared with analytical results based of the experimental parameters (blue line).

## 5. Conclusion

To conclude, several fundamental features of the phase evolution of Peregrine soliton or NLS rational solitons have been experimentally observed in this work. We have provided the first experimental evidence in hydrodynamics of the clear $\pi$ phase shifts that exist between the central peak and the continuous background. In the case of PS, the central part is in antiphase with respect to the continuous background. By contrast, for the second-order rational soliton, the central peak is in phase with the continuous background. In nonlinear optics, we have provided additional experimental confirmation of the $\pi$ phase-shift that exists between the central part of the pulse and the continuous background. A careful analysis of a series of longitudinal measurements has enabled us to also confirm the phase excursion experienced by the central peak with respect to the continuous background. Due to the interdisciplinary character of the approach, this study may emphasize a wide range of applications in other nonlinear dispersive media. It could be extended to other governing equation such as the Lugiato-Lefever equation that governs the nonlinear dynamics in resonators where pulsating solution may also exist [42]. It also raises new questions regarding the influence of the initial phase profile of the excitation in the one-dimensional focusing NLSE [43].

Those various results have been obtained taking advantage of relatively basic analysis tools. They could be easily extended to other physical domains such as the analysis of the time series recorded in multicomponent plasma [8]. Note that more advanced analysis techniques have now become available [36, 44]. Our longitudinal optical measurements are intrinsically discrete as they are based on destructive cut-back methods [32] or on the use of different sections of fibers. Use of a recirculating loop could be an efficient alternative [45]. The progress in optical characterization has now made possible continuous longitudinal measurements using the coherent analysis of the Rayleigh back reflected wave [41]. All these possibilities may stimulate a renewed interest in the study of phase shifts that may emerge upon propagation of nonlinear waves [12, 46].

# Acknowledgements

We acknowledge the support of the Institut Universitaire de France (IUF), the French Investissements d'Avenir program and the Agence Nationale de la Recherche (EIPHI Graduate School ANR-17-EURE-0002; ISITE-BFC ANR-15-IDEX-0003 Projects Bright and Nextlight).